\documentclass[12pt]{article}

\usepackage{verbatim}
\usepackage{amsmath}
\usepackage{amssymb}
\usepackage{graphicx}
\usepackage{cite}
\usepackage{subfig}
\usepackage{setspace}
\usepackage{url}
\usepackage[percent]{overpic}
\usepackage{slashed}
\usepackage{authblk}
\usepackage{color}

\usepackage{xspace}

\usepackage{fullpage}
\usepackage{hyperref}

\def\ie{{\it i.e.}}
\def\eg{{\it e.g.}}

\def\to{\rightarrow}

\title{The ATLAS $Z+$MET Excess in the MSSM}
\author{M. Cahill-Rowley$^a$}
\author{J.L. Hewett$^a$}
\author{A. Ismail$^{b,c}$}
\author{T.G. Rizzo$^a$}

\affil{$^a$~SLAC National Accelerator Laboratory, Menlo Park, CA, USA\footnote{mrowley, hewett, rizzo@slac.stanford.edu}}
\affil{$^b$~Argonne National Laboratory, Argonne, IL, USA\footnote{aismail@anl.gov}}
\affil{$^c$~University of Illinois at Chicago, Chicago, IL, USA}

\begin{document}

\rightline{\vbox{\halign{&#\hfil\cr
&SLAC-PUB-16308\cr
}}}


{\let\newpage\relax\maketitle}

\begin{abstract}

We demonstrate that the $3\sigma$ excess observed by ATLAS in the $Z+$MET channel can be explained within the context of
the MSSM.  Using the freedom inherent in the pMSSM, we perform a detailed analysis of the parameter space and find a scenario that describes the excess while simultaneously 
complying with all other search constraints from the Run I data at 7 and 8 TeV, including the $Z+$MET analysis by CMS.  We generate
a small sample of simplified models, using promising models from our existing pMSSM sample as seeds, and study their properties.
The successful region
is described by the production of $1^{st}/2^{nd}$ generation squark pairs, followed by their decay into a bino-like neutralino which in turn decays into
a Higgsino-like LSP triplet by emitting a $Z$  boson, {\it i.e.}, $\tilde q\to\tilde B\to\tilde h$ with $\tilde q = \tilde Q_L,\tilde u_R,$ or $\tilde d_R$.  The sweet spot for the sparticle spectrum is found to have squark
masses in the 500-750 GeV range, with bino masses near 350 GeV with a mass splitting of 150-200 GeV with the Higgsino LSP.   If this excess holds, then this scenario predicts that a signal will be observed in the $0l+$jets and/or $1l+$jets searches in the early operations of Run II.

\end{abstract}

\newpage

\section{Introduction}

It is well-known that physics beyond the Standard Model (SM) must exist in order to address a number of outstanding questions such as
the nature of dark matter, the generation of neutrino masses, the origin of the observed baryon asymmetry and 
the solution to the hierarchy problem -- all of which remain unanswered. The nature of this new physics is presently
mysterious: Not only is its form unknown, so is the energy scale at which it will first be revealed.  Although constrained by data from Run I
at the LHC,  dark matter searches, and flavor physics observables, Supersymmetry (SUSY)\cite{SUSYrefs} remains the leading theoretical framework to address at least some of these important puzzles. 
However, Supersymmetry has so far been frustratingly elusive at the LHC, with numerous searches setting strong constraints on the
simplest SUSY scenarios\cite{SUSYsearches}.  Nonetheless, the continual exploration of the SUSY parameter space remains mandatory, with
missing transverse energy (MET) based searches at the LHC continuing to be the most promising avenue for discovery.

Along these lines, the ATLAS experiment recently announced\cite{ATLASsignal} the observation of a $3\sigma$ excess in one of their Run I SUSY search 
channels, $Z+$MET with $\geq 2j$, while a similar analysis by CMS\cite{CMSnonsignal} observed a result consistent with the expected 
SM background.   Importantly, the detailed nature of the cuts employed by these two experiments in this channel  
are sufficiently different, as we will discuss below, so that the apparent null result from CMS does not necessarily exclude the possibility of 
a signal being observed by ATLAS.  However, an explanation of this potential signal within Supersymmetry remains challenging,
since any proposed scenario must also satisfy the constraints imposed by the plethora of ATLAS and CMS 
searches\cite{SUSYsearches}. Nonetheless, a few new physics scenarios have been proposed\cite{ZMETpapers} that could give rise to the observed ATLAS excess
with varying degrees of success.  In this work, we suggest a natural Supersymmetric scenario, based on the pMSSM, which comfortably explains the ATLAS excess in the $Z+$MET channel while evading all other searches.     

Given the simple nature of the search channel, and the apparent rate of the excess, several features are 
clearly necessary for a Supersymmetric model to provide a successful description of the data. 
Since the $Z$-boson is observed in the dilepton mode, the signal rate demands a strong production cross section, implying the
production of relatively light gluinos or squarks which then decay to an intermediate 
state accompanied by jets. This intermediate, apparently neutral state, {\it e.g.}, a neutralino, then decays via the 
emission of a $Z$ plus the lightest Supersymmetric particle (LSP), which produces the MET in an R-parity 
conserving scenario{\footnote {In the analysis considered here, we will assume the LSP to be the lightest neutralino.}}. 
However, it is likely that such a spectrum would be easily excluded by, \eg, the 0l, $2-6$ jets+MET searches 
if the jets from the hadronic decay of the $Z$ were sufficiently hard. Clearly the details of the SUSY spectrum in such a scenario, 
in particular the relative masses and compositions of the sparticles, 
are highly constrained by multiple requirements and finding the right `balance' 
presents a significant model building challenge.

To perform this study, it is necessary to incorporate a detailed analysis of the available SUSY parameter space that remains viable after
the Run I data at 7 and 8 TeV; there is no better way to accomplish this than to employ the 19-parameter 
p(henomenological)MSSM\cite{pMSSM} which we have already studied in detail elsewhere\cite{us}.  In particular, this recent work contains a large sample of
pMSSM models that are presently allowed, providing a viable playground for exploration.  As will be discussed in detail below, an 
examination of these models reveals an intriguing scenario that describes the excess, while complying with all the constraints.
Specifically,  the $1^{st}/2^{nd}$ squarks, $\tilde Q_L, \tilde u_R$ or $\tilde d_R$ are identified as the leading 
candidates for the objects that initiate the ATLAS $Z+$MET signal via a cascade decay.  Once the other search constraints are taken
into account, the primary production of gluinos in 
this role are found to yield too small of an event rate to explain the signal.
The $1^{st}/2^{nd}$ generation squark scenario benefits from having both of the first two generations of squarks being 
produced simultaneously, as they are assumed to be degenerate in the pMSSM framework, yielding a large enough production
rate.  Whereas, within our pMSSM model sample, the $3^{rd}$  generation 
squarks are also too highly constrained by specialized searches to play the role of 
the strong initiator of this signal. Within our successful scenario, the $1^{st}/2^{nd}$ generation squarks decay to a mostly bino-like neutralino 
which then subsequently decays to a Higgsino-like 
LSP triplet by emitting a $Z$-boson. The masses and splittings dictated by this spectrum control the overall production rates for the different sparticles, the hardness  
of the jets and leptons, and the branching fractions for the intermediate neutralino decaying to the three light Higgsino states.
We note that considering only a single set of squarks presents a somewhat simplified picture and that other states (such as $\tilde t$ and $\tilde b$) may also contribute to the total signal, albeit in a secondary capacity.  In the analysis below we use the successful models contained in our existing  pMSSM sample as seeds to generate a small sample of
simplified models that describe the excess while remaining consistent with the many other LHC searches.  We then study the detailed properties of these simplified models and discuss the Run II analyses that can be used to elucidate this scenario more fully, or exclude it from further consideration.

\section{Analysis}

The pMSSM\cite{pMSSM} is the most general version of the R-parity conserving MSSM subjected to the guiding principles of CP-conservation, Minimal Flavor Violation, and degenerate 1$^{st}$ and 2$^{nd}$ generation squark masses.  Imposing these criteria reduces the number of free parameters in the MSSM to 19 (assuming a neutralino LSP):  $m_{\tilde Q_{L1,2}}$,  $m_{\tilde Q_3}$, $m_{\tilde u_{R1,2}/\tilde d_{R1,2}/\tilde t_R/\tilde b_R}$, $m_{\tilde L_{L1,2}}$, $m_{\tilde L_3}$, $m_{\tilde e_{R1,2}/\tilde\tau_R}$, $M_{1,2,3}$, $\mu$, $A_{t,b,\tau}$, 
$M_A$, and $\tan\beta$.  In our previous work\cite{us}, we generated a large set of models (with a `model' describing a point in the 19-dimensional parameter space) by randomly scanning the parameter space, setting the upper limit on the scan at 4 TeV for the dimensionful parameters and taking $\tan\beta=1-60$.  The 4 TeV upper bound was chosen to facilitate collider studies at the 7,8 and 13,14 TeV LHC.  We subjected these models to a global data set of collider, flavor, precision, dark matter and theoretical constraints.
In particular, we have examined this model sample in light of the SUSY search results from Run I of the LHC\cite{us}, subjecting them to roughly 40 separate analyses performed by the experiments at 7 and 8 TeV.  The result is a sample of approximately 125k models that remain viable at the end of Run I, including many models with light squarks and gluinos, providing an ample playground for further studies. 

Here, we investigate our pMSSM model sample to determine whether a region of the parameter space could adequately describe the excess in the $Z+$MET channel observed by ATLAS. For each model, we created SUSY production samples using Pythia 6.4 \cite{Sjostrand:2006za} for event generation and PGS \cite{PGS} for detector simulation, normalizing to NLO cross sections from Prospino \cite{Beenakker:1996ch}. The details of this procedure are the same as for our previous pMSSM studies \cite{us}. We then applied the cuts for the on-$Z$ region of the ATLAS search for final states containing a pair of opposite-sign dileptons, jets and MET \cite{ATLASsignal} to these simulated SUSY samples. In particular, events were required to have at least two leptons with $p_T > 25, 10$ GeV respectively, two jets with $p_T > 35$ GeV, and MET $>$ 225 GeV. The two hardest leptons were required to form a same-flavor opposite sign pair with invariant mass $m_{\ell\ell} = m_Z\ \pm\ 10$ GeV, and the scalar sum $H_T$ of their transverse momenta and the transverse momenta of all jets with $p_T > 35$ GeV was required to be at least 600 GeV. We also imposed standard cuts on the rapidity and isolation of jets and leptons, and required an angular separation $\Delta \phi (j, \mathrm{MET}) > 0.4$ between each of the two leading jets and the missing transverse momentum, as described by ATLAS.

\begin{figure}[htbp]
\centering
\includegraphics[width=0.6\textwidth]{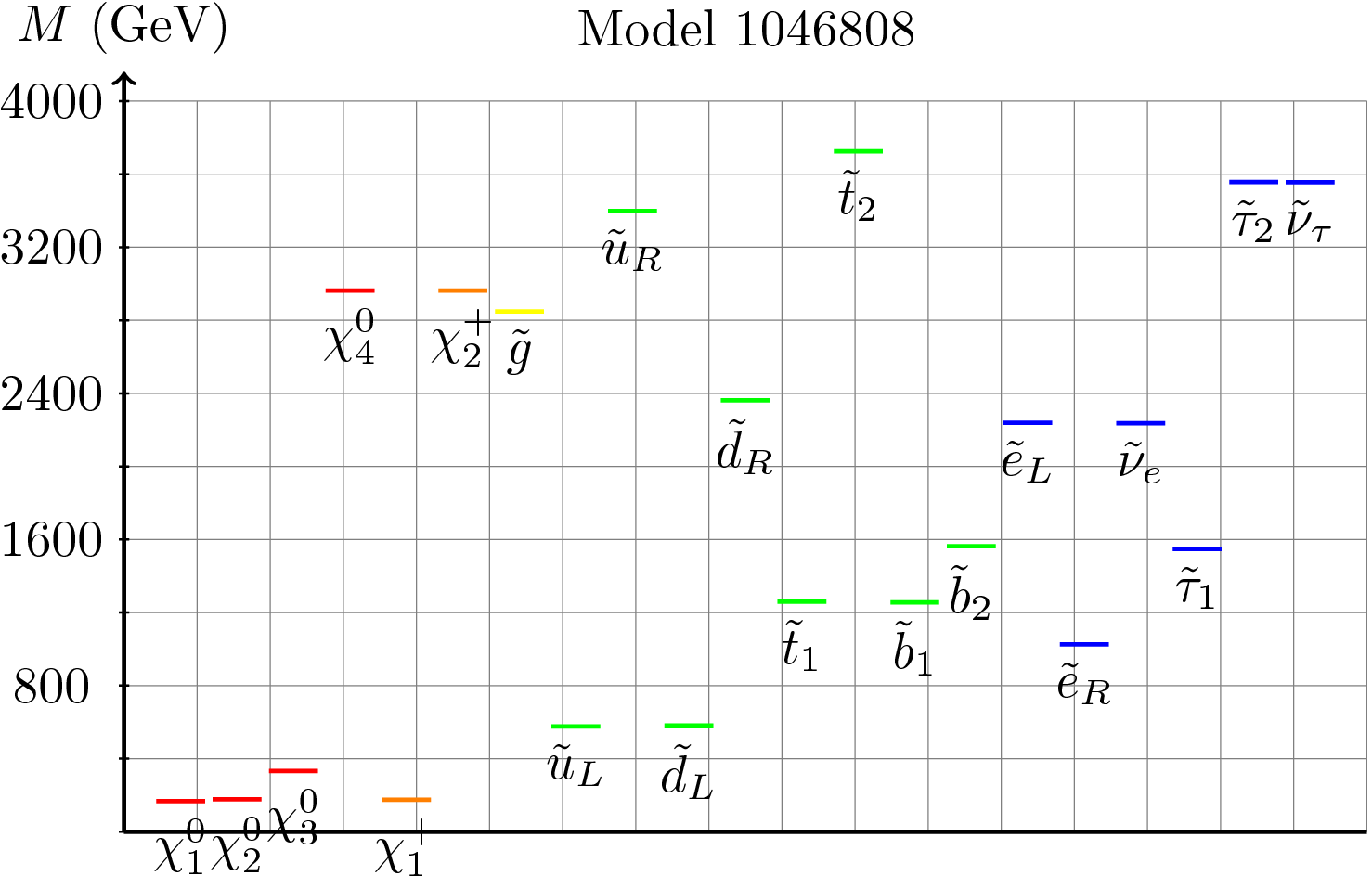}
\caption{Sparticle spectrum for a representative pMSSM model that reproduces the ATLAS $Z+$MET excess.}
\label{fig:spspect}
\end{figure}

Several of our pMSSM models predict significant numbers of events passing the ATLAS $Z+$MET cuts. The sparticle spectrum for such a representative pMSSM model is shown in Fig.~\ref{fig:spspect}. We observe a common pattern in these models, with light-flavor squarks decaying through gaugino cascades producing $Z$-bosons. The direct decay of the squarks to the lightest neutralino is usually suppressed by weak couplings to a Higgsino-like LSP (or alternatively a wino-like LSP if the squark is right-handed). Additionally, we find that these models tend to predict an observable excess in jets + MET searches, creating some tension with the null results in other Run I LHC SUSY searches that we have considered previously \cite{us}. In particular, it is challenging to reproduce the ATLAS $Z+$MET excess in the leptonic channel while simultaneously satisfying bounds from the ATLAS jets + MET search \cite{TheATLAScollaboration:2013fha}. However, the pMSSM models from our sample that reproduce the $Z+$MET excess often predict
a jets + MET event rate that is near the boundary of the existing limits. In particular, the number of events in the jets + MET search signal regions is typically reduced because the squarks decay mainly to a heavier bino-like neutralino rather than directly to the LSP.  

Encouraged by these results, we are motivated to consider points with similar spectra to these promising pMSSM models, which may predict a significant number of $Z+$MET events while fully evading constraints from the other LHC SUSY searches. Given the results of our pMSSM analysis, we focus on simplified spectra with the dominant decay pattern $\tilde{q} \to \tilde{B} \to \tilde{h}$, where $\tilde{q}$ is a light-flavor squark $\tilde{Q}_L$, $\tilde{u}_R$ or $\tilde{d}_R$.
Starting with a seed point taken from one of our successful pMSSM models, we vary the 3 most relevant 
Lagrangian parameters ($m_{\tilde Q_{L1,2}}$, $\mu$, and $M_1$) in a grid around the region of interest. This corresponds to adjusting the most relevant 
physical sparticle masses, specifically $m_{\tilde\chi{_1^0}}$, $m_{\tilde\chi{_3^0}}$, and the squark masses 
$m_{\tilde q}$. Note that neither the sign of $\mu$ or the value of $\tan\beta$ are varied, as this would modify the 
details of the Higgsino spectrum yet leave the gaugino branching fractions mostly unaltered.\footnote{Except for cases where some of the Higgsinos are kinematically inaccessible to decays of the bino-like $\tilde{\chi}_3^0$.} We expect the remaining pMSSM parameters to have a negligible impact on the simplified model phenomenology, as long as they are sufficiently heavy.  We thus leave their values as given in the pMSSM seed model, with the exception of $A_t$ which is adjusted to produce the observed value of the Higgs mass within theoretical and experimental uncertainties. In all cases, the squark masses not being scanned are set above 2 TeV. 

To produce this grid, we scan $\mu$ between 100 GeV and 254 GeV (with the lower limit set by LEP constraints and upper limit set by kinematic considerations), with 22 GeV steps.  Note that given this small value of $\mu$ we would
expect these models to exhibit low values of fine-tuning from this source. 
We then scan the \textit{physical} squark mass $m_{\tilde Q_L} \simeq m(\tilde{u}_L)\simeq m(\tilde{d}_L)$ between 350 GeV or $\mu + 150$ GeV (whichever is larger) and 900 GeV in increments of 35 GeV \footnote{Since we don't know the physical squark mass before spectrum generation, we estimate the soft mass required to give the desired physical mass. While approximate, this estimation is easily accurate enough to ensure that our scan grid is covering the region of interest.}. Finally, we scan $M_1$ between $\mu + 100$ GeV and $m_{\tilde Q_L}$ with 25 GeV increments. We employ the same procedure (with slightly different scan ranges noted in Table~\ref{GridScan}) to construct two additional grids, one each for  $\tilde u_R/\tilde c_R$ and $\tilde d_R/\tilde s_R$.

\begin{figure}[htbp]
\centering
\includegraphics[width=0.8\textwidth,height=0.9\textwidth]{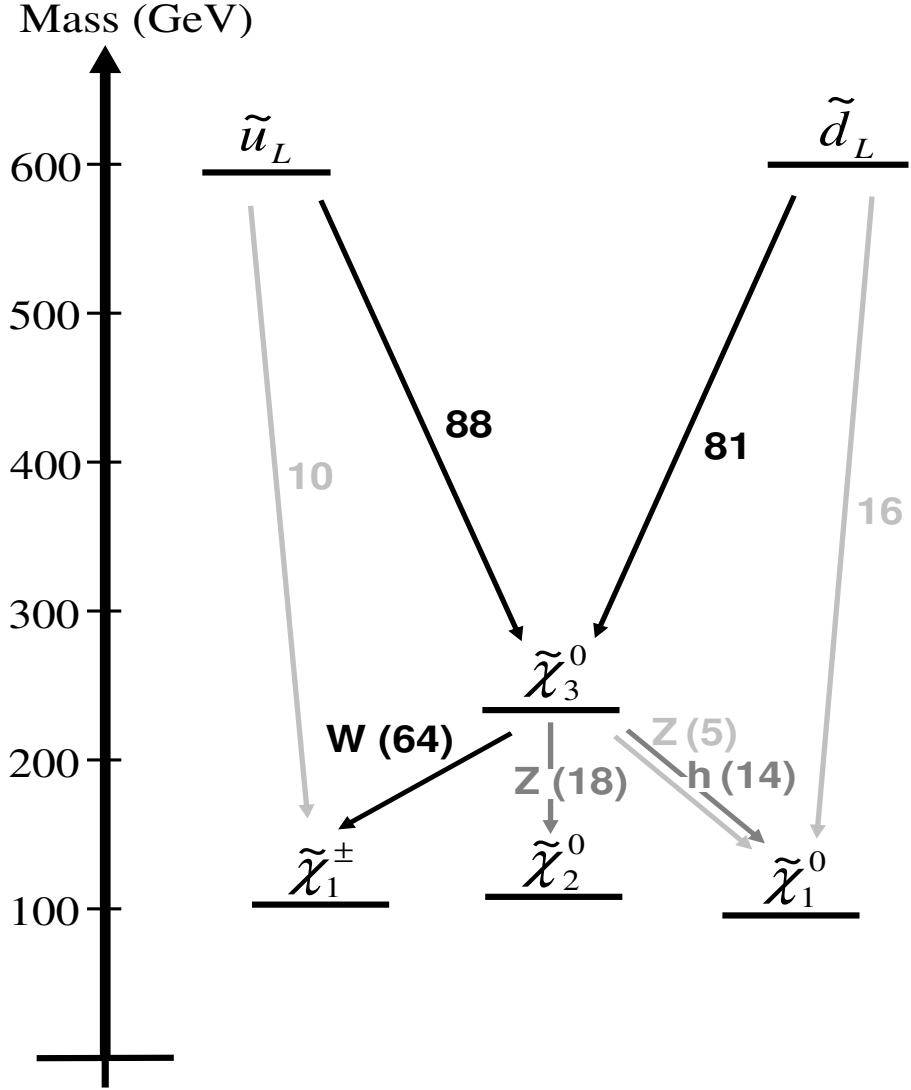}
\caption{Spectrum and decay patterns for a model in the $\tilde Q_L$ grid, which predicts 21 events in the ATLAS 20 fb$^{-1}$  
$Z$+MET analysis and is consistent with all other implemented searches. Numbers indicate the branching fraction in percent for each decay mode (only branching fractions larger than 5\% are shown for simplicity).}
\label{fig:decay_patterns}
\end{figure}

Figure~\ref{fig:decay_patterns} shows the relevant spectrum and branching fractions for one of our grid points (from the $\tilde Q_L$ grid), which predicts 21 events in the ATLAS Z + MET search and is consistent with all other searches; this model is illustrative of the typical decay patterns for scenarios that reproduce the excess. The key features to note are the large branching fractions for squarks decaying to $\tilde{\chi}_3^0$, and the multiple possible decays of $\tilde{\chi}_3^0$, about a quarter of which result in $Z$ boson production.

\begin{table}[htbp]
\centering
\begin{tabular}{|l|l|l|l|} \hline\hline
Grid & $\mu$ (22 GeV steps) & $M_1$ (25 GeV steps)& $m_{\tilde Q_L}$ (35 GeV steps)  \\
\hline
$\tilde{Q}_L$ & 100 GeV - 254 GeV  & $\mu$ + 100 GeV - $m_{\tilde Q_L}$ & MAX(350 GeV, $\mu$ + 150 GeV) - 900 GeV \\
$\tilde{u}_R$ & 100 GeV - 254 GeV & $\mu$ + 100 GeV - $m_{\tilde Q_L}$ & MAX(300 GeV, $\mu$ + 150 GeV) - 800 GeV \\
$\tilde{d}_R$ & 100 GeV - 254 GeV & $\mu$ + 100 GeV - $m_{\tilde Q_L}$ & MAX(250 GeV, $\mu$ + 150 GeV) - 700 GeV \\

\hline\hline
\end{tabular}
\caption{Scan ranges for the 3 variable parameters in each of the 3 grid scans described in the text.}
\label{GridScan}
\end{table}

\section{Results}

We now examine the results of our scan over the simplified pMSSM spectra.
As noted above, the strongest restrictions on the parameter space arise from the null results of other LHC SUSY searches, which are generally in tension with our goal of producing a large signal rate in the ATLAS 20 fb$^{-1}$ $Z$+MET analysis. Clearly, we require that a successful model point produce $\sim 15-20$ signal events for the ATLAS 20 fb$^{-1}$  
$Z$+MET analysis. In addition, we also require the point to simultaneously satisfy the limit from the corresponding CMS search with different 
selection criteria. Finally, a successful model point must satisfy all of the null ATLAS search results in other channels. In particular, it is clear that both the ATLAS 1l+jets search (arising in the spectra we consider from the heavy bino decay producing a $W$ instead of a $Z$) and the 0l, 2-6 jets search (when $W\,,Z$, or the Higgs are produced and decay hadronically) will also be important in determining the detailed nature of a successful parameter space point. The impact of these other searches will be discussed in more 
detail below.

\begin{figure}[htbp]
\centering
\includegraphics[width=0.3\textwidth]{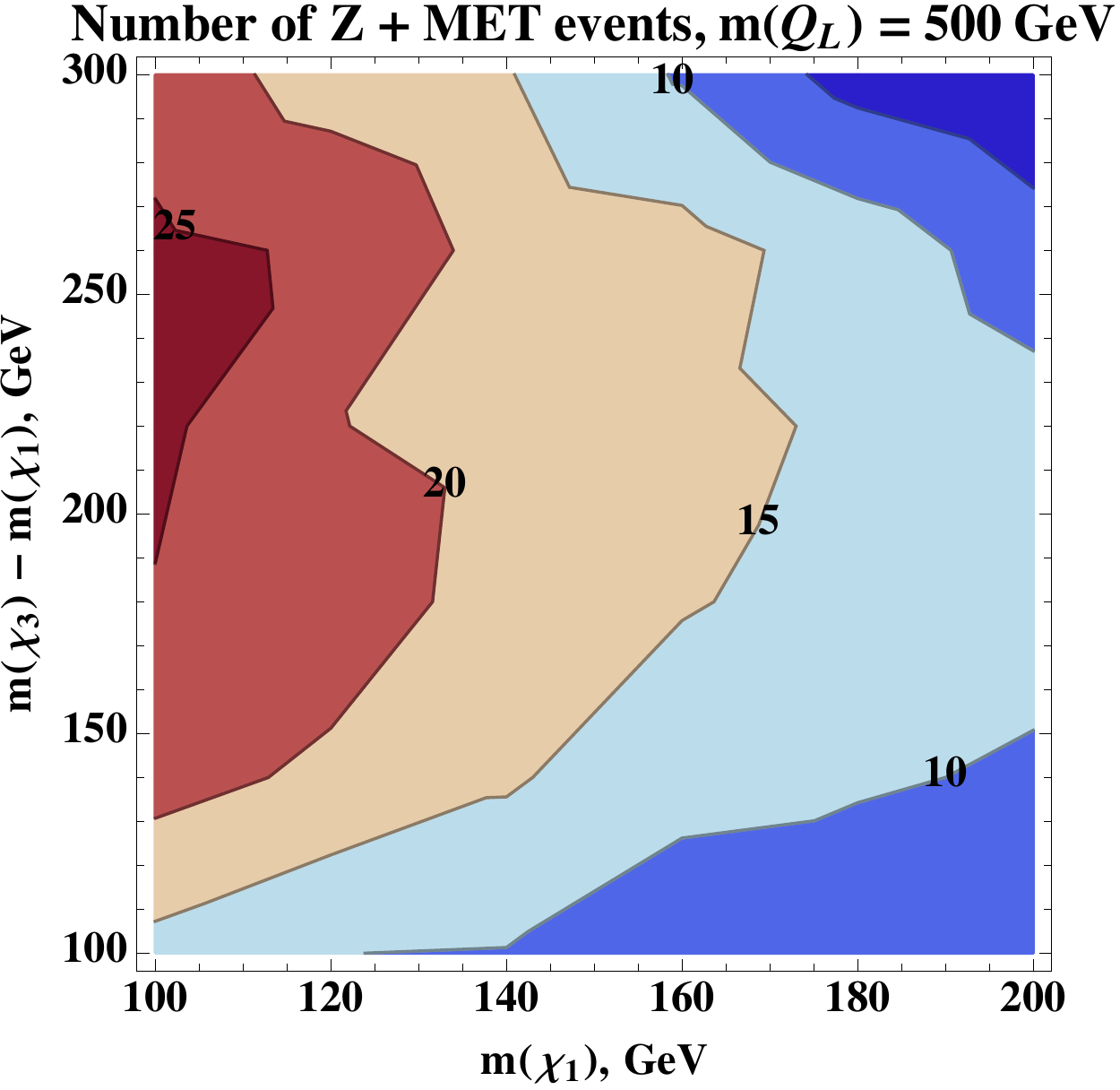}
\includegraphics[width=0.3\textwidth]{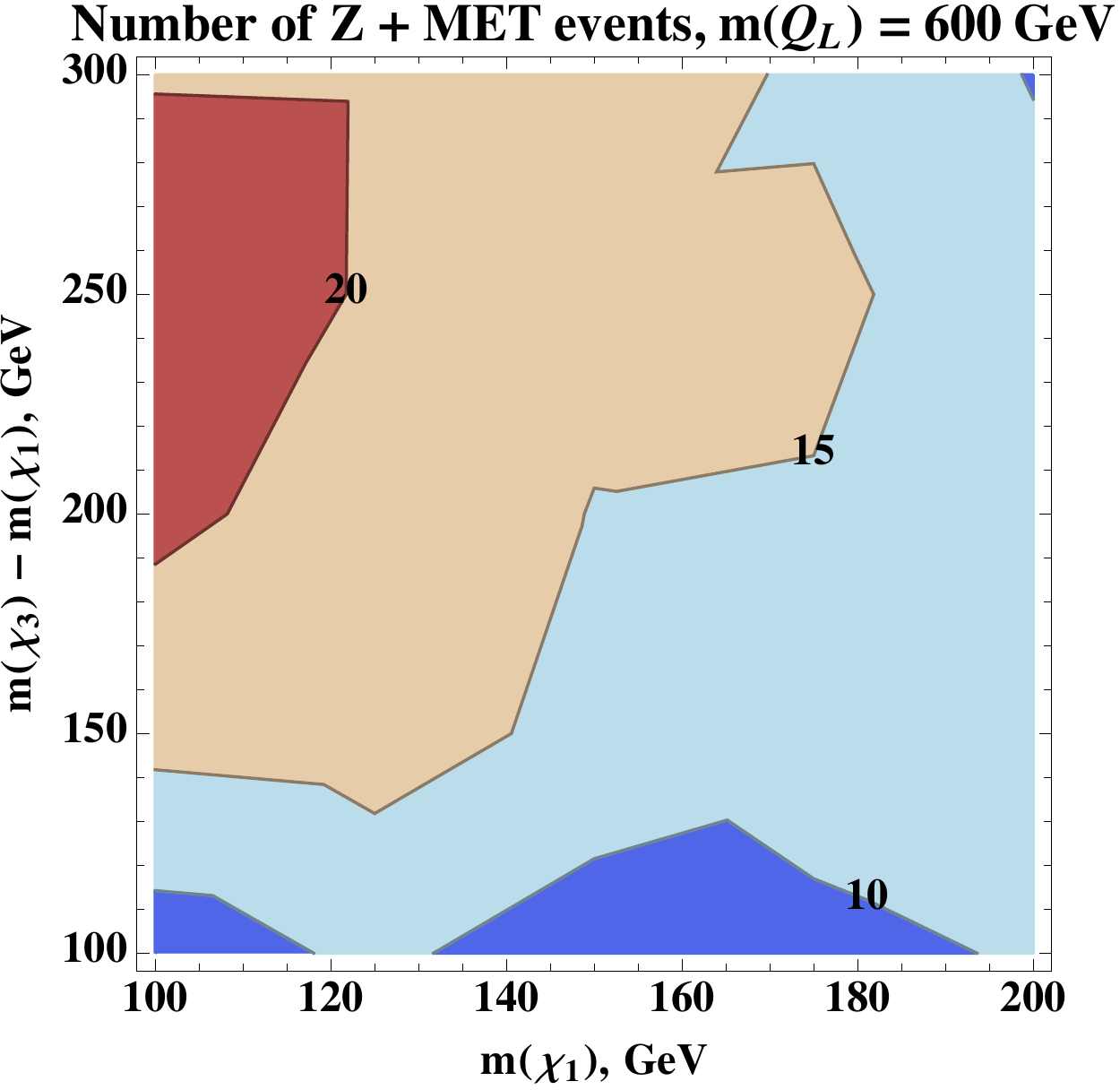}
\includegraphics[width=0.3\textwidth]{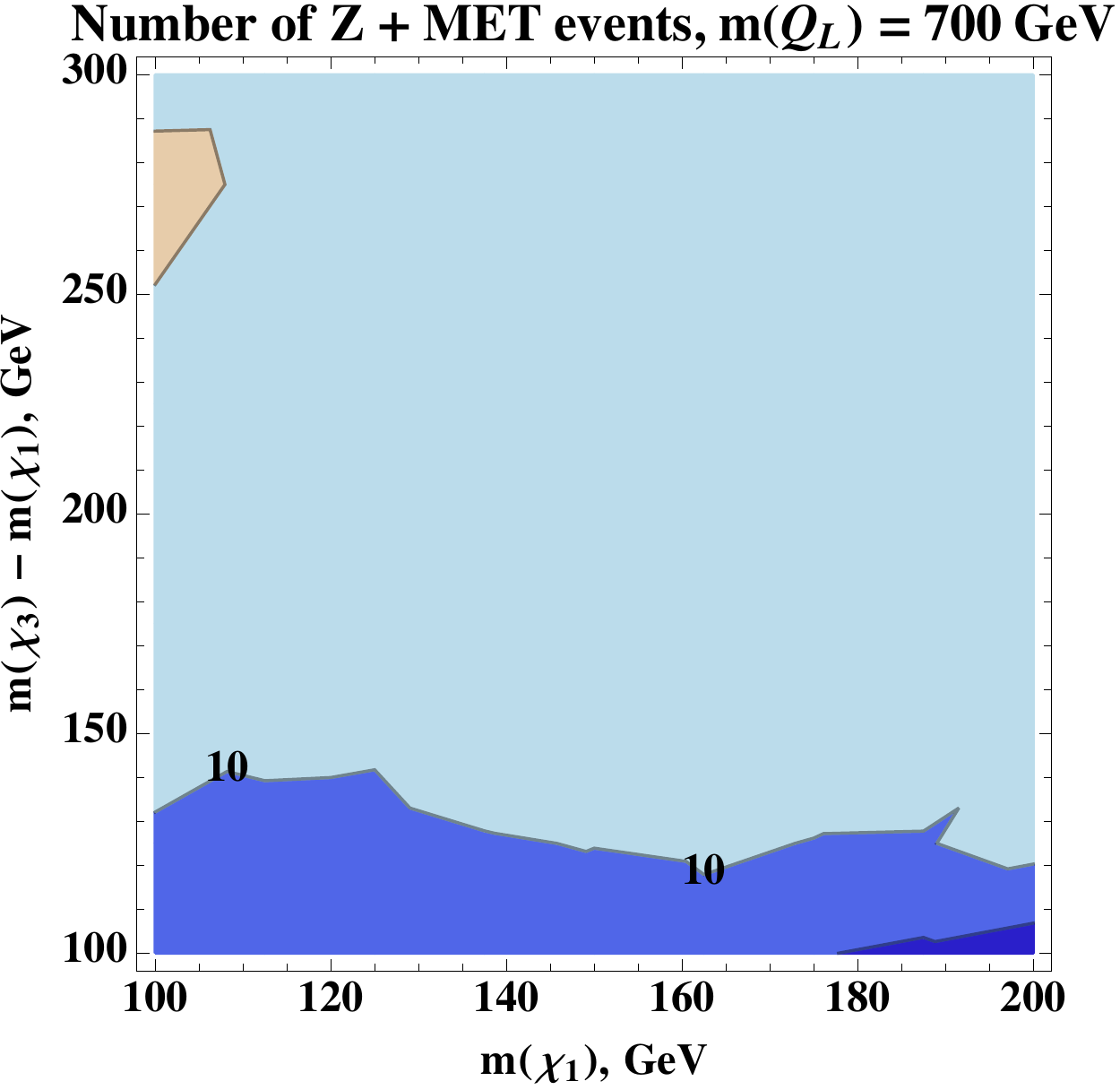}
\includegraphics[width=0.3\textwidth]{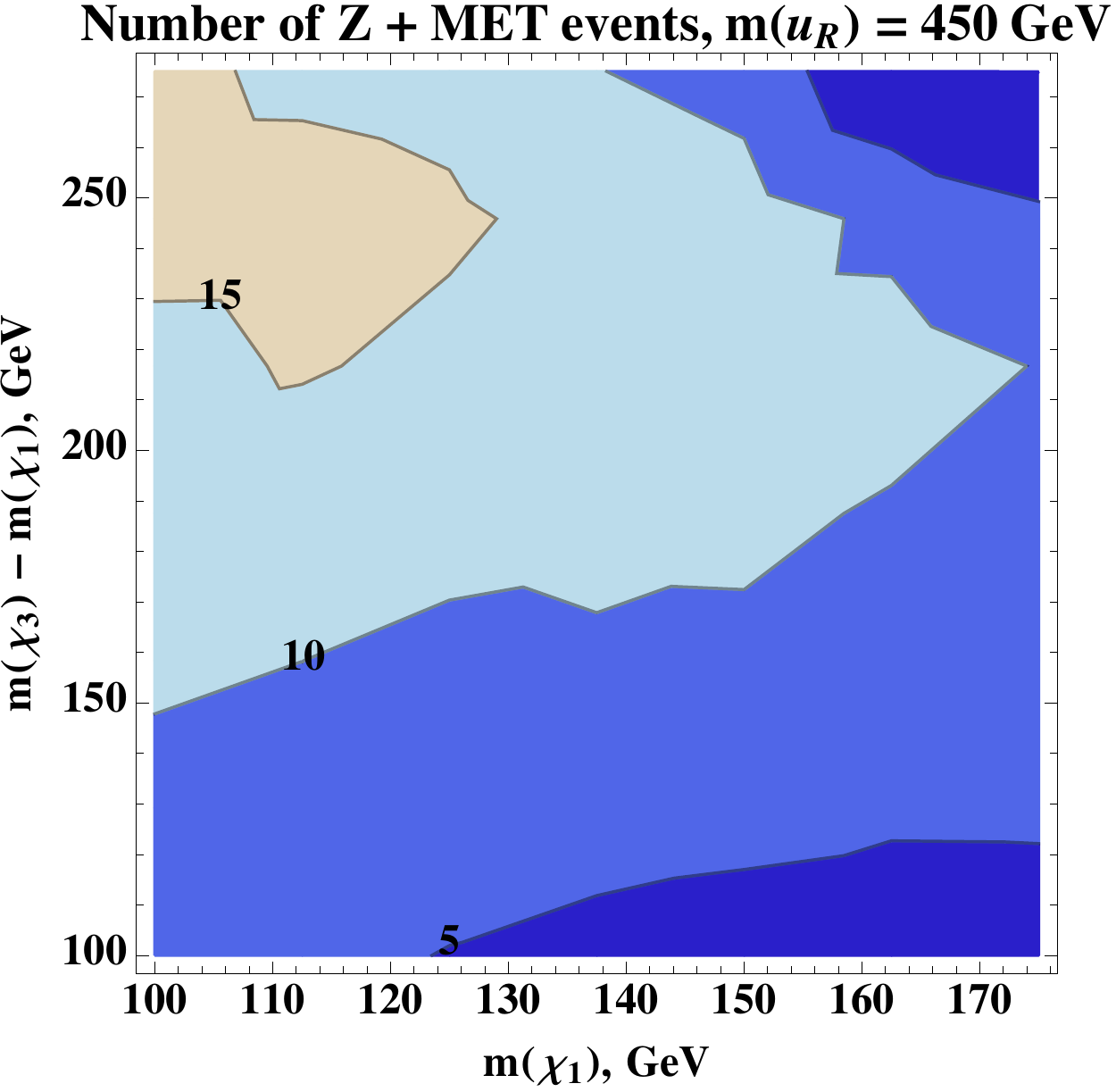}
\includegraphics[width=0.3\textwidth]{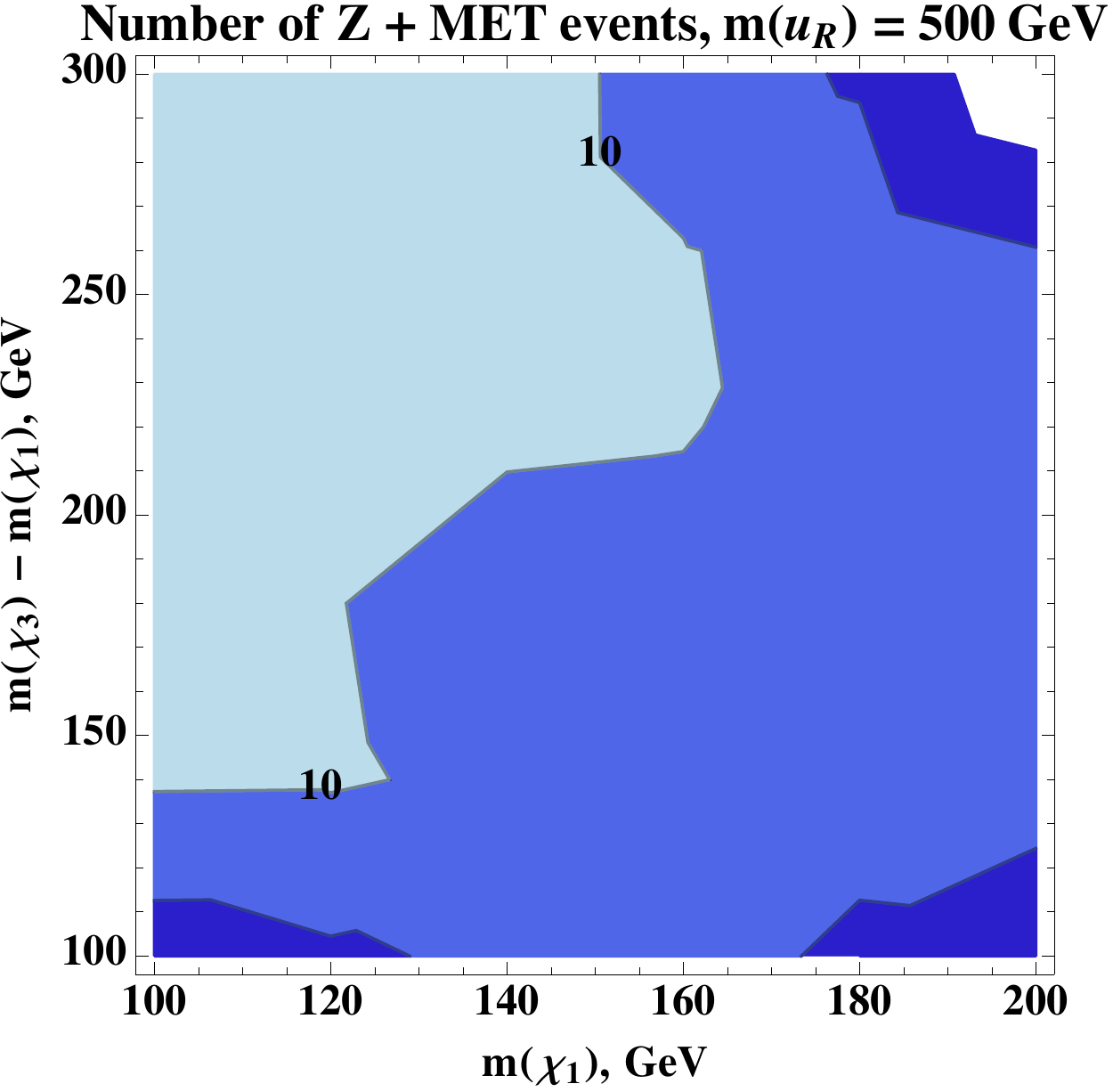}
\includegraphics[width=0.3\textwidth]{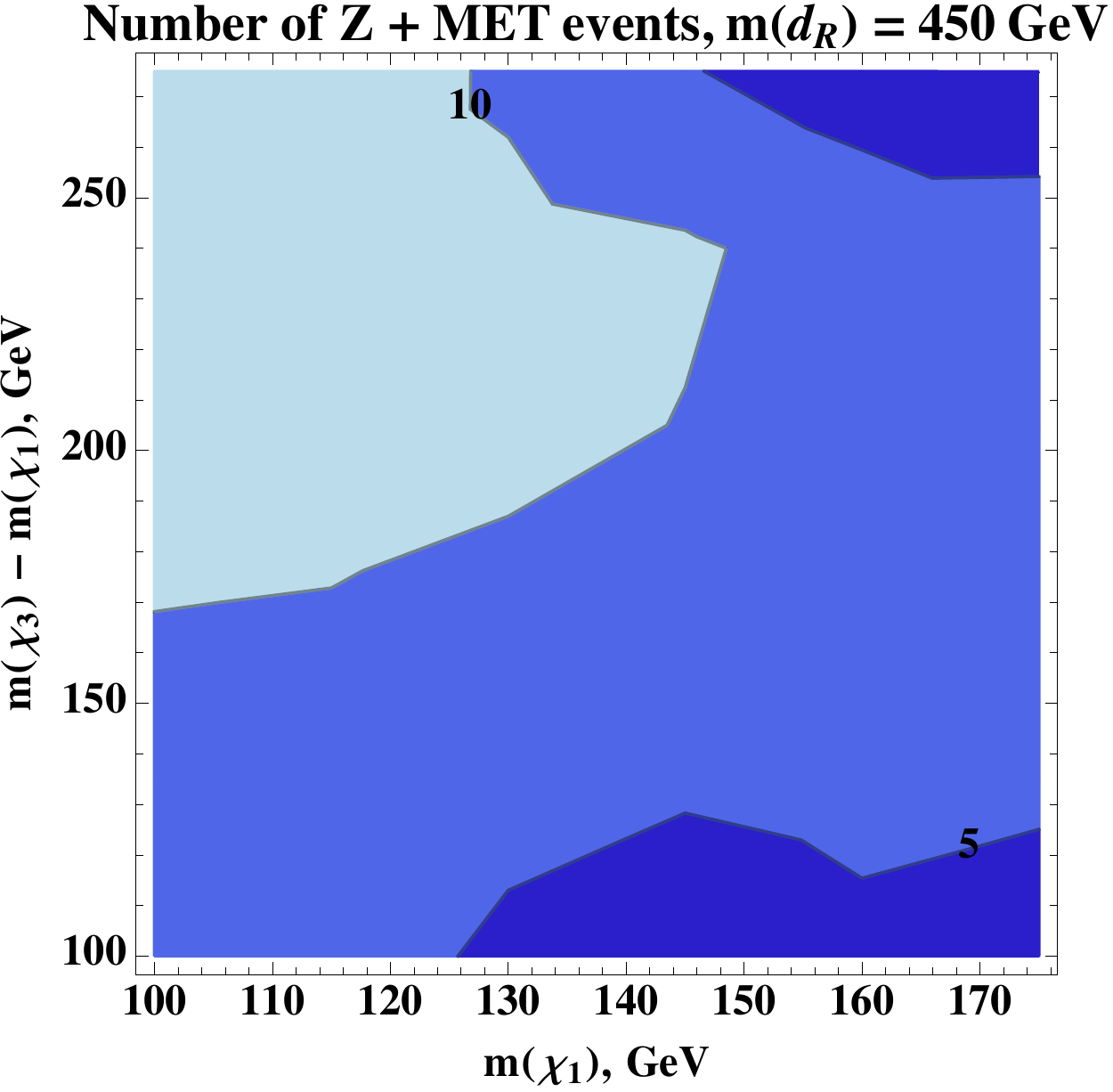}
\caption{Signal event rate contours for the ATLAS $Z+$MET analysis in the $\chi_3^0-\chi_1^0$ mass difference and $\chi_1^0$ mass plane.
The top three panels correspond to the case of $\tilde Q_L=500\,, 600\,, 700$ GeV from left to right, while the bottom panels are for
$\tilde u_R=450\,, 500$ GeV and $\tilde d_R=450$ GeV, left to right.}
\label{fig:zjmetevents}
\end{figure}

To get an initial handle on the preferred parameter regions, Fig.~\ref{fig:zjmetevents} shows the LSP mass versus the mass splitting between the intermediate bino-like $\tilde \chi_3^0$ and the LSP for fixed values of the squark masses in the 450-700 GeV range {\it {before}} imposing any additional constraints.\footnote{Note that we have interpolated between grid points, smoothing out modest fluctuations in event yields.} The various 
colored regions show the anticipated ATLAS $Z$+MET analysis event yields and, as we would naively expect, we see that 
lighter squarks will generally lead to larger signal rates due to their significantly larger production cross sections. Also, we 
see that the Higgsino-like LSP prefers to be relatively light, below $\sim 180-190$ GeV. Perhaps, most interestingly, 
we observe that the most favored range for the electroweakino mass splitting lies above $\sim 150$ GeV. This might be 
counter-intuitive since we would naively expect that a mass splitting, $\Delta m_{31}$, in the range of  $\sim 90-125$ GeV would be 
most desirable, since decays to the Higgs in this region would be kinematically forbidden, thereby increasing the 
branching fraction for decays through the $Z$. Clearly, in all cases we see that the largest signal rates are obtained when the 
$\chi_3^0$ is kinematically allowed to decay through both the $Z$ and the Higgs, due to the increased visibility of the decay products. Of course the preferred range 
of $\Delta m_{31}$ is somewhat sensitive to the nature of the parent squark. In the $\tilde Q_L$ case, a value of $\Delta m_{31} \sim 150-200$ 
GeV is preferred while for $\tilde u_R(\tilde d_R)$ this value is significantly larger $\Delta m_{31} \sim 200-300(250-350)$ GeV (as we will see more clearly below) in obvious 
correlation with the production cross sections, \ie, the parent squark with the largest (smallest) production cross section 
prefers the smallest (largest) corresponding value of $\Delta m_{31}$.

\begin{figure}[htbp]
\centering
\includegraphics[width=1.00\textwidth]{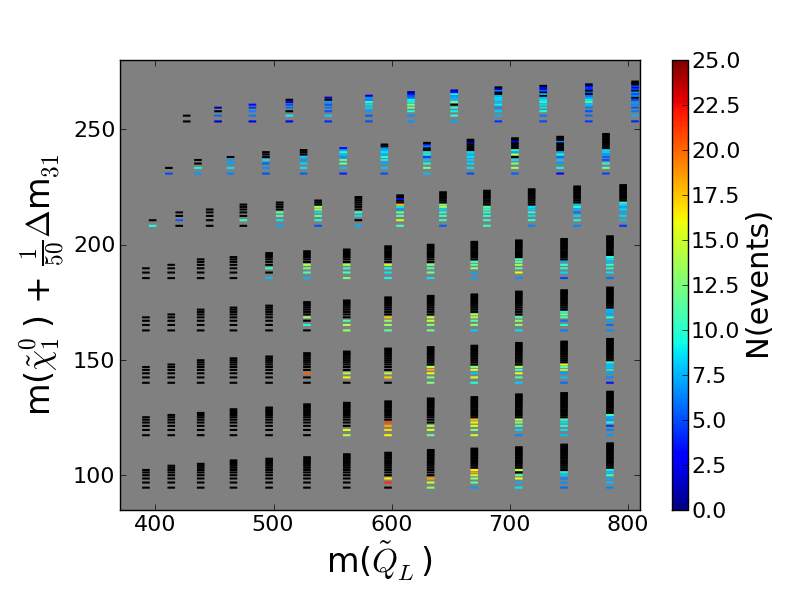}
\caption{Results from the simplified spectra scan for a parent $Q_L$-squark in the $\tilde Q_L$ and $\chi_1^0+ (1/50)\Delta m_{31}$ mass plane.  The vertical bars represent the coarse grid in our scan, with the value of the mass splitting $\Delta m_{31}$ increasing for successively higher slices of the bar.  The color code
indicates the predicted event rate for the ATLAS 20 fb$^{-1}$ $Z+$MET channel.  Black slices in a vertical bar correspond to points excluded by any of the simulated searches.}
\label{fig:scans1}
\end{figure}

\begin{figure}[htbp]
\centering
\includegraphics[width=1.00\textwidth]{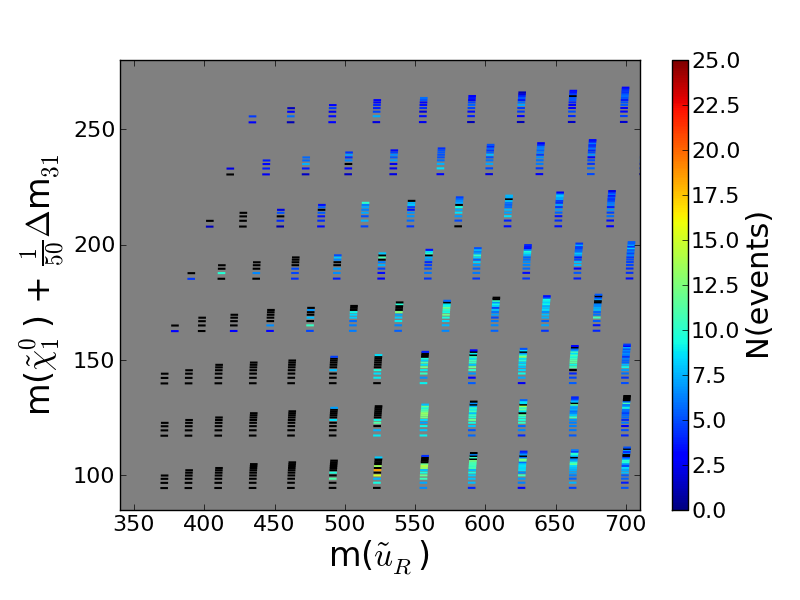}
\caption{Results from the simplified spectra scan for a parent $u_R$-squark in the $\tilde u_R$ and $\chi_1^0+ (1/50)\Delta m_{31}$ mass plane.  The vertical bars represent the coarse grid in our scan, with the value of the mass splitting $\Delta m_{31}$ increasing for successively higher slices of the bar.  The color code
indicates the predicted event rate for the ATLAS 20 fb$^{-1}$ $Z+$MET channel.  Black slices in a vertical bar correspond to points excluded by any of the simulated searches.}
\label{fig:scans2}
\end{figure}

\begin{figure}[htbp]
\centering
\includegraphics[width=1.00\textwidth]{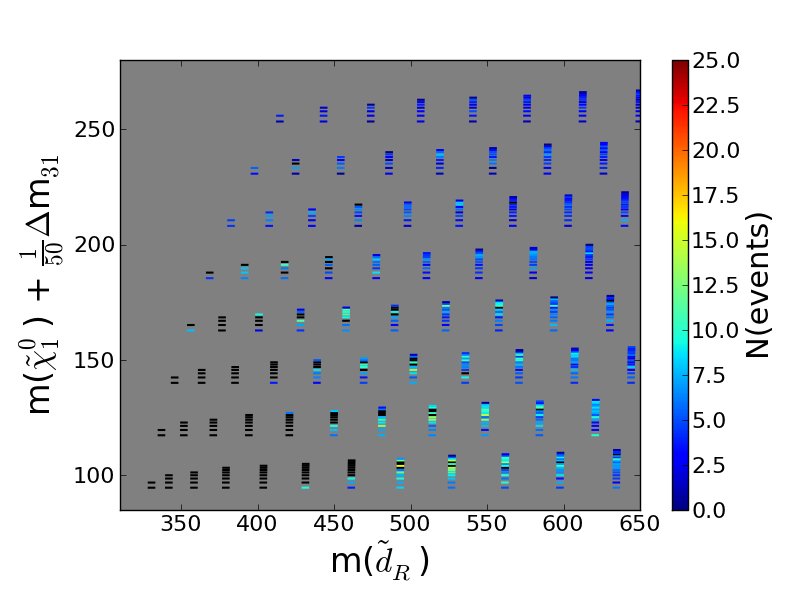}
\caption{Results from the simplified spectra scan for a parent $d_R$-squark in the $\tilde d_R$ and $\chi_1^0+ (1/50)\Delta m_{31}$ mass plane.  The vertical bars represent the coarse grid in our scan, with the value of the mass splitting $\Delta m_{31}$ increasing for successively higher slices of the bar.  The color code
indicates the predicted event rate for the ATLAS 20 fb$^{-1}$ $Z+$MET channel.  Black slices in a vertical bar correspond to points excluded by any of the simulated searches.}
\label{fig:scans3}
\end{figure}

Figures \ref{fig:scans1}, \ref{fig:scans2} and \ref{fig:scans3} present
the results from our scans of the simplified pMSSM spectra for each parent squark type, 
where the vertical bars represent the scanned regions in our somewhat 
coarsely spaced grid.  The location of the vertical bars is set by the approximate mass of the 
parent squark and the LSP mass. The lowest slice 
of each of the vertical bars corresponds to the smallest value of the bino-LSP mass splitting, $\Delta m_{31}$, scaled by a factor of 0.02,
while each successive higher slice (going up the bar) corresponds to increasing this value. The color code 
indicates the number of events predicted for the ATLAS 20 fb$^{-1}$ $Z$+MET analysis, with yellow to red tones indicating a higher
event rate in agreement with the observed rate. Black regions represent points which 
are excluded by the corresponding CMS $Z$+MET search, or by any of the $\sim$ 40 null SUSY searches described in~\cite{us}. In the $\tilde Q_L$ grid, the 
0l+jets channel results gives by far the most important constraints, while for $\tilde u_R$, and particularly $\tilde d_R$, other searches also play an important role.
The results are seen to be quite different for the three parent squark 
cases we consider.  In particular, we see that the $\tilde Q_L$ case provides the best fit to the excess. However, even in this case
the most successful points are close to the black excluded regions, indicating that the other LHC SUSY searches are providing important constraints on this scenario. We also find the strong constraints at larger values of $\Delta m_{31}$, due to larger contributions to the 0l+jets channel when this splitting is too large.  In general, we expect the 0l+jets rate to place strong restrictions on the $\tilde Q_L$ scenario because of either a large production rate for relatively light squark masses or because the rate is still reasonably large for heavier masses where the jets from the decay of the $W\,, Z$ and Higgs bosons appearing in the bino to LSP transitions are becoming sufficiently hard to pass the 0l+jets cuts. For surviving points explaining the Z + MET excess, we find that low LSP masses and moderate values of $\Delta m_{31}$ are preferred, and that the production
rate falls off too quickly for squark masses above $\sim 800$ GeV to generate a sufficient number of events.
In the case of a $\tilde u_R$ parent, both the favored region and the region excluded by the other searches 
are smaller (a simple consequence of the lower production cross section) and, overall, lower signal rates are obtained.  
These same features are seen to be further emphasized for the case where the parent squark is a $\tilde d_R$.  In all cases, we find that the
sweet spot for describing the excess is in the region where the parent squark is $500-700$ GeV with a LSP mass of $100-200$ GeV and a bino-LSP mass splitting of $100-250$ GeV.

\begin{figure}[htbp]
\centering
\includegraphics[width=0.45\textwidth]{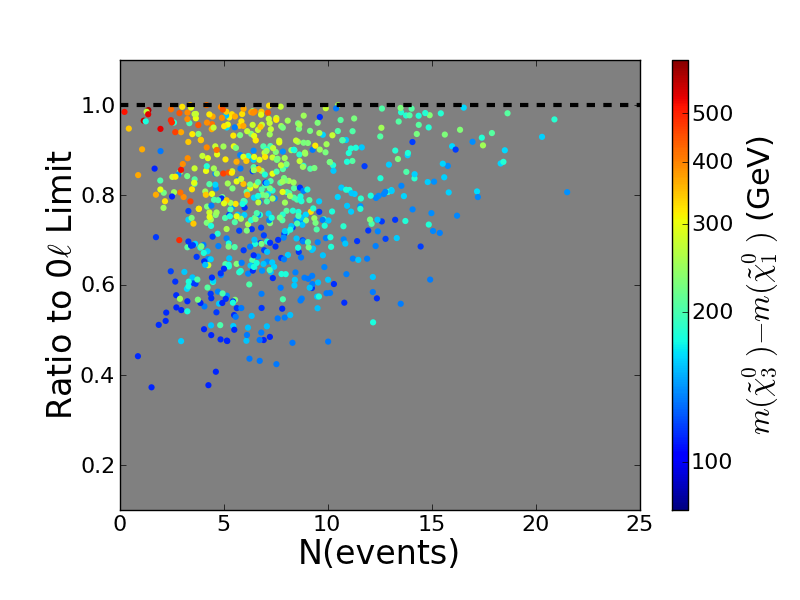}
\includegraphics[width=0.45\textwidth]{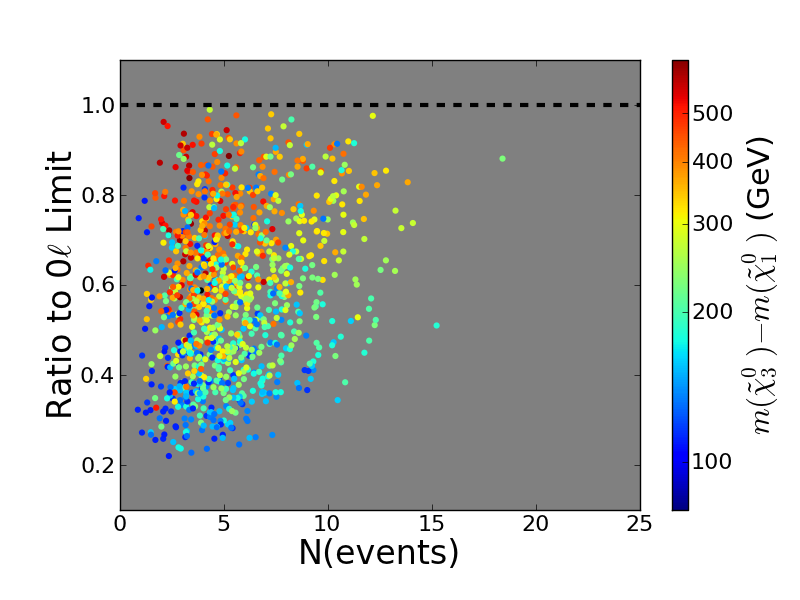}
\includegraphics[width=0.45\textwidth]{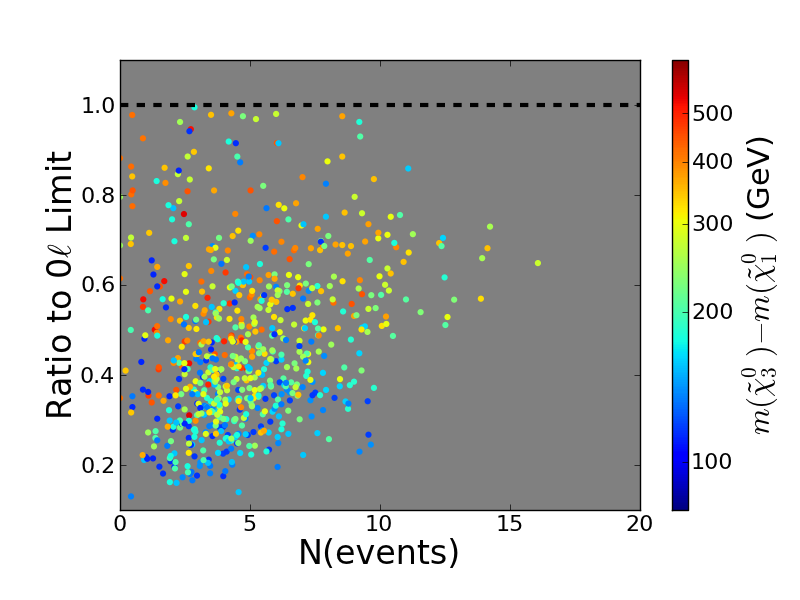}
\caption{Ratio of the predicted number of events for models in our simplified grid scan to the ATLAS 95\% C.L. event limit for the 0l+jets
channel, $R_{0l}$, as a function of the number of signal events for the ATLAS 20 fb$^{-1}$ $Z+$MET search.  The color code corresponds to the value of the 
$\chi_3^0-\chi_1^0$ mass splitting.  The top-left, top-right, and bottom panels correspond to the three grid scans, with $\tilde Q_L$,
$\tilde u_R$, and $\tilde d_R$ parent squarks, respectively.}
\label{fig:0l}
\end{figure}

For each parent squark type, we next examine the impact of the 0l,1l+jets and CMS $Z+$MET channels.  Here, we only study the
set of models from our simplified grid scan that are consistent with the constraints in these channels. For each of these analyses, we compute the expected number of events in each signal region, and show the ratio of the expected number of events to the 95\% C.L. event exclusion limit for the most 
important signal region for that channel (\textit{i.e.}, the signal region with the largest value of this ratio).  For example, a ratio of 0.5 indicates that the model predicts 
1/2 as many events as are allowed by the relevant null search result at $95\%$ C.L. Figure \ref{fig:0l} displays these results for the 
case of the 0l+jets ATLAS search, {\it i.e.}, the event rate ratio $R_{0l}$, as a function of the number of predicted events for the ATLAS 20 fb$^{-1}$ $Z$+MET analysis for all three parent squark types.  The color code indicates the value of $\Delta m_{31}$.  The top-left, top-right, and bottom panels correspond to the parent squarks $\tilde Q_L$,
$\tilde u_R$, and $\tilde d_R$, respectively. In the $\tilde Q_L$ parent case, we see 
that many of the models lie close to the 0l+jets exclusion boundary. In particular, 
we see that for model points with at least 15 ATLAS 20 fb$^{-1}$ $Z+$MET signal events, the values of $R_{0l}$ lie in the range 0.6-1. Generally 
an increase in the number of ATLAS 20 fb$^{-1}$ $Z+$MET signal events corresponds to a larger value of $R_{0l}$, so that at some point consistency
with the 0l+jets search prevents larger signal rates from being obtained. Also, we see that as  
$\Delta m_{31}$ decreases for a fixed signal rate, the points are farther away from the 0l+jets exclusion boundary since the jets produced by $W\,, Z$ and Higgs 
decays are becoming correspondingly softer.  Considering the $\tilde u_R$ parent case, we find that the model points are a 
bit further away from the 0l+jets boundary (due to the smaller production cross section), but we also find, correspondingly, 
fewer models that produce a significant signal in the $Z+$MET analysis. This trend continues for the case of the 
$\tilde d_R$ parent.

\begin{figure}[htbp]
\centering
\includegraphics[width=0.45\textwidth]{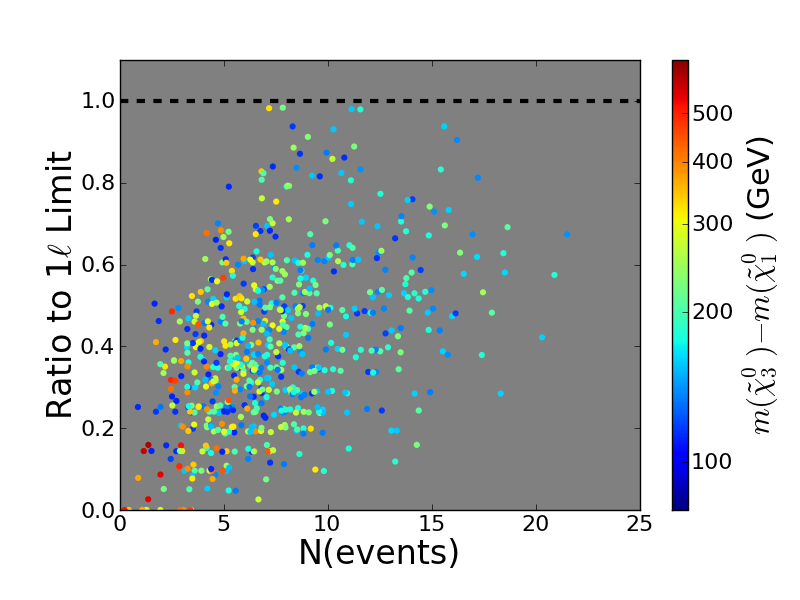}
\includegraphics[width=0.45\textwidth]{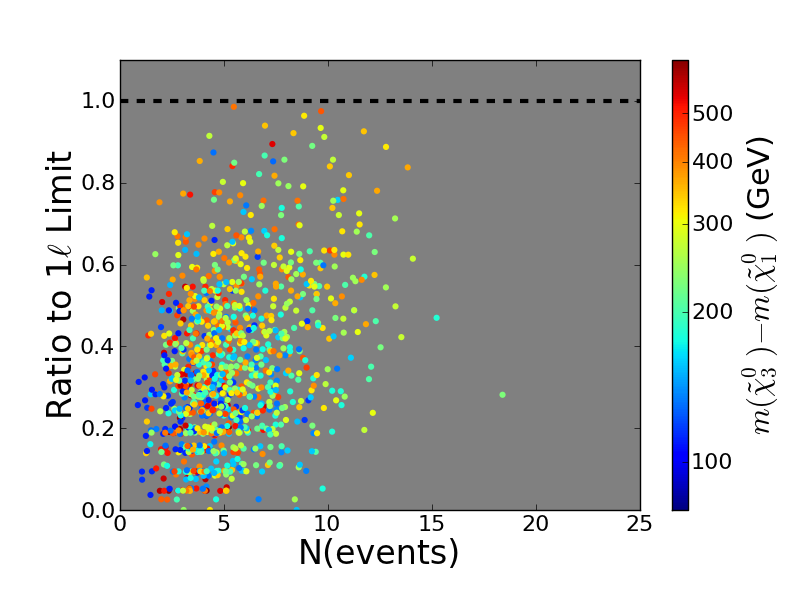}
\includegraphics[width=0.45\textwidth]{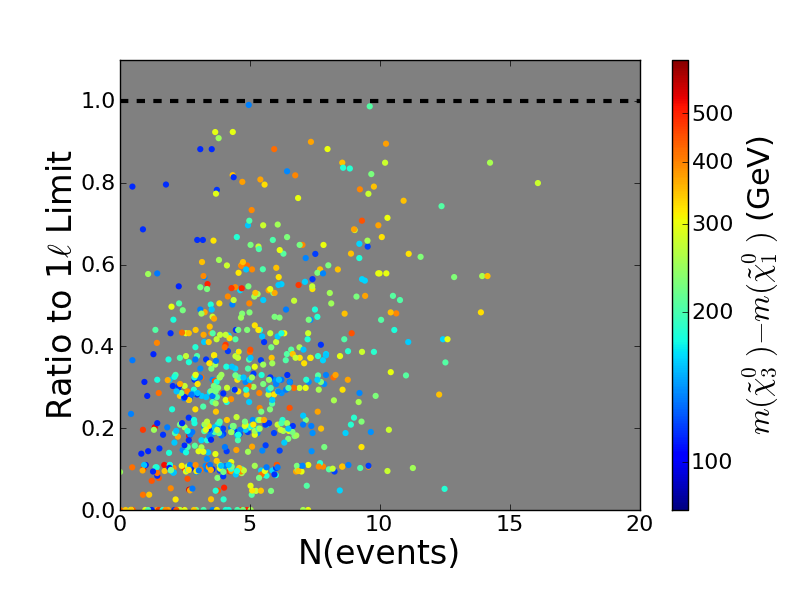}
\caption{Ratio of the predicted number of events for models in our simplified grid scan to the ATLAS 95\% C.L. event limit for the 1l+jets
channel, $R_{1l}$, as a function of the number of signal events for the ATLAS 20 fb$^{-1}$ $Z+$MET search.  The color code corresponds to the value of the 
$\chi_3^0-\chi_1^0$ mass splitting.  The top-left, top-right, and bottom panels correspond to the parent squark cases of $\tilde Q_L$,
$\tilde u_R$, and $\tilde d_R$, respectively.}
\label{fig:1l}
\end{figure}

\begin{figure}[htbp]
\centering
\includegraphics[width=0.45\textwidth]{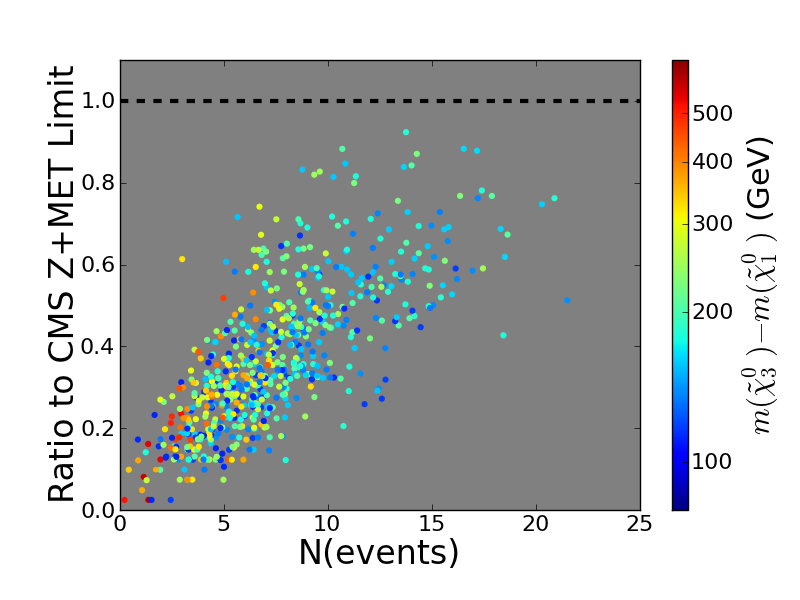}
\includegraphics[width=0.45\textwidth]{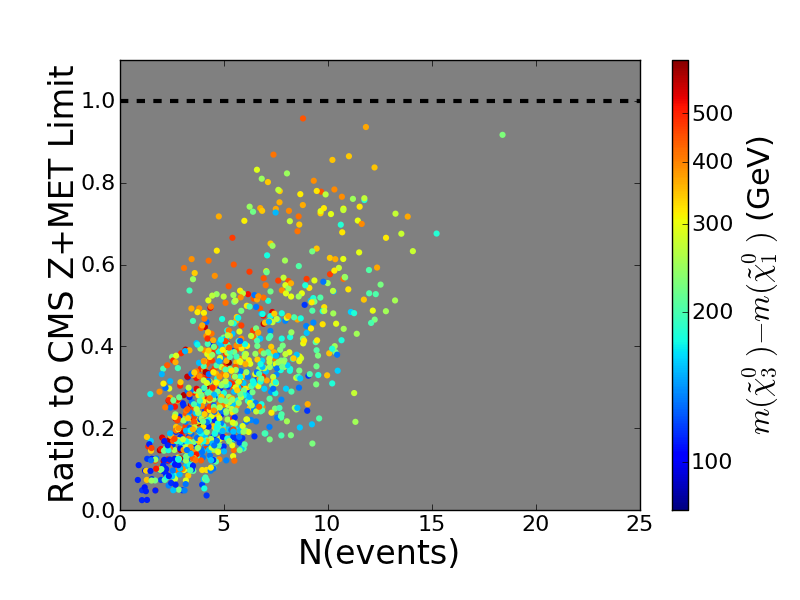}
\includegraphics[width=0.45\textwidth]{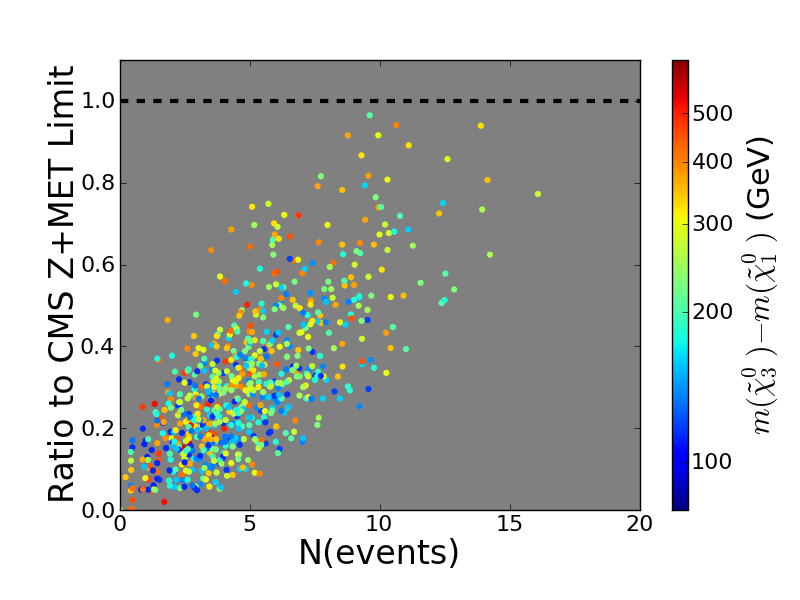}
\caption{Ratio of the predicted number of events for models in our simplified grid scan to the CMS 95\% C.L. event limit for the $Z+$MET
channel, $R_{CMS}$, as a function of the number of signal events for the ATLAS 20 fb$^{-1}$ $Z+$MET search.  The color code corresponds to the value of the 
$\chi_3^0-\chi_1^0$ mass splitting.  The top-left, top-right, and bottom panels correspond to the parent squark cases of $\tilde Q_L$,
$\tilde u_R$, and $\tilde d_R$, respectively.}
\label{fig:2lCMS}
\end{figure}

Figure \ref{fig:1l} displays our results for the case of the 1l+jets ATLAS search where  
the y-axis now shows the event rate ratio $R_{1l}$. For all three squark parent cases we see that the models tend to mostly lie 
reasonably far away from the exclusion boundary for this search, implying that it has little impact on 
shaping the parameter region for successful models. In fact, we find that few models are excluded by the 1l+jets analysis after the other 
null search results have been applied. Figure \ref{fig:2lCMS} shows the results for the case of the CMS $Z+$MET analysis, 
expressed as the ratio $R_{CMS}$; clearly for all squark parents there is a rough linear correlation between the 
value of $R_{CMS}$ and the number of predicted ATLAS 20 fb$^{-1}$ $Z$+MET signal events. From this one might expect that requiring 
$R_{CMS} \leq 1$ cuts off the corresponding ATLAS signal. However, this region is 
already restricted by the 0l+jets ATLAS search, with the result that the CMS $Z+$MET analysis has 
only a small additional impact on our model selection beyond the effect of the 0l+jets ATLAS search.

In addition to the 0l,1l+jets and CMS $Z+$MET searches, other ATLAS searches which are less clearly targeted for these types of models can still have an important impact on the allowed parameter space, especially for the $\tilde u_R$ and $\tilde d_R$ grids, where the 0l search is less dominant. In particular, the ATLAS 3l gaugino search~\cite{ATLAS:2012uks}, 4l search~\cite{ATLAS:2012hmt}, and same-sign dilepton search~\cite{ATLAS:2012sna} all make unique contributions to the combined exclusion region. Since the 3l and 4l searches are targeted at electroweak production, it is unsurprising that they are particularly sensitive to the lower mass regions allowed for the right-handed squarks, particularly $\tilde d_R$.  

It is worth a short discussion to compare the event selection between the ATLAS and CMS $Z+$MET analyses. While both searches select events with a leptonic Z, at least two jets, and missing energy, the 600 GeV $H_T$ cut of the ATLAS search is highly effective at reducing Drell-Yan background, leaving $t\bar{t}$ as the dominant background process. The CMS analysis considers multiple search regions with missing energy bins to gain increased sensitivity, but even in the highest bin, requiring MET $> 300$ GeV, Drell-Yan production is still the most significant background. For comparison, the ATLAS analysis imposes the tight $H_T$ cut stated above and simply cuts on missing energy, MET $> 225$ GeV. As a result of these cut choices, the overlap between the ATLAS and CMS $Z+$MET search regions is small \cite{haastalk}.

\begin{figure}[htbp]
\centering
\includegraphics[width=0.45\textwidth]{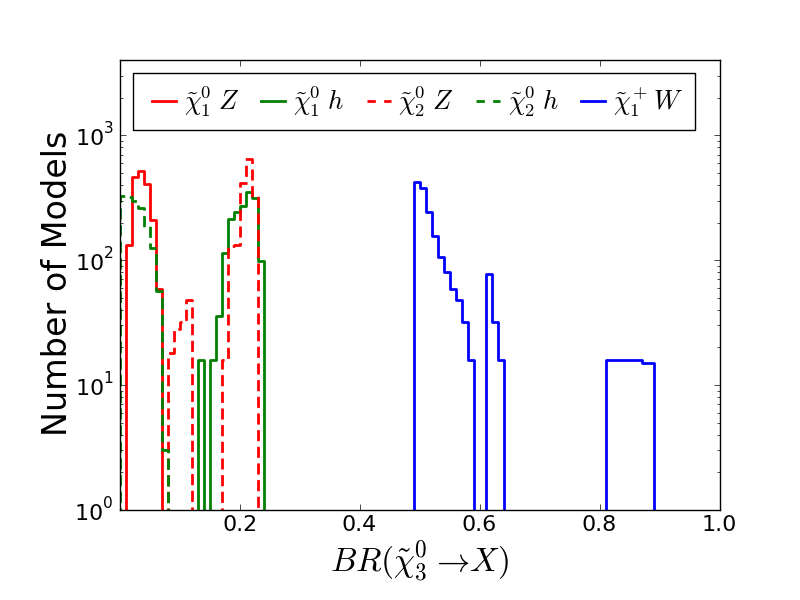}
\includegraphics[width=0.45\textwidth]{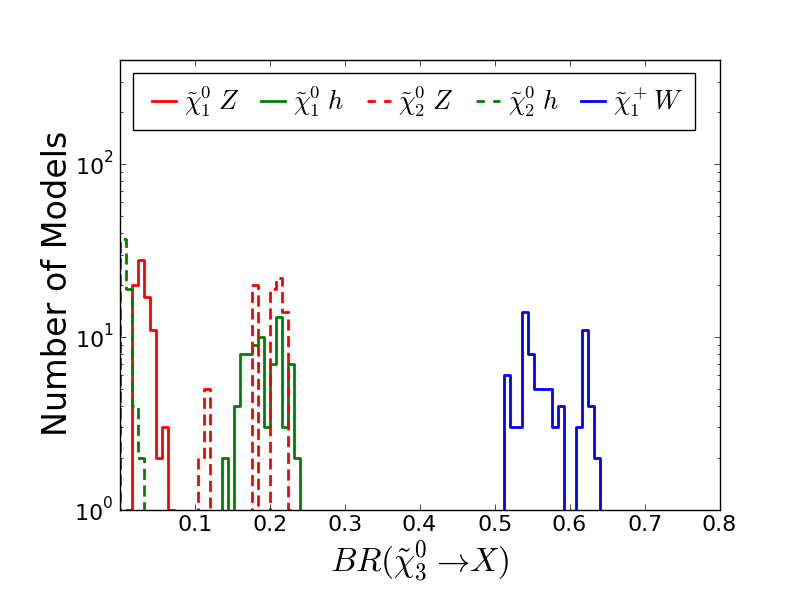}
\caption{Distributions of the branching fractions for the decay modes, as indicated, of the $\chi_3^0$ intermediate state for the full set of models from our $\tilde{Q}_L$ grid 
scan (left panel) and for models predicting 12 or more events in the ATLAS 20 fb$^{-1}$ $Z+$MET search while remaining consistent with all simulated searches, including the ATLAS 0l,1l+jets and CMS $Z+$MET searches (right panel).}
\label{fig:bfs}
\end{figure}

As discussed above, our successful models kinematically allow the decay of the intermediate bino-like state into the lighter 
Higgsinos by $W\, ,Z$ and Higgs boson emission. Apart from phase space considerations, these relative branching fractions are 
controlled by the bino and Higgsino content of the gauginos.\footnote{We essentially treat the winos as being decoupled with a 
correspondingly large value of $M_2$.} Since $\tan \beta$ is being held fixed in our grid scans, the bino and Higgsino content of 
these states are only being regulated (at tree-level) by the values of $M_1$ and $\mu$. Clearly as the mass splitting between these states, 
$\Delta m_{31}$, increases, the purity of each state increases. If the intermediate state were to be 
{\it pure} bino, then its decay via either the $W$ or $Z$-boson would be forbidden (as these channels require both the initial and final 
states to have a nonzero Higgsino content), while decays through the Higgs would remain allowed (as this is controlled 
instead by the product of the bino and Higgsino content of both the initial and final states).  Figure \ref{fig:bfs} shows the distributions
for the $\chi_3^0$ branching fractions for the $\tilde Q_L$ parent squark case for all the models in the initial grid, 
as well as after applying constraints from the CMS $Z+$MET search and null results in other ATLAS channels and requiring the point to predict 12 or more events in the ATLAS 20 fb$^{-1}$ $Z+$MET search. Here we see several things: ($i$) the typical $W$-boson branching fraction is rather large, although models with the largest values for this branching fraction are unable to satisfy the constraints applied to the right panel. 
($ii$) In both panels, the $Z$-boson branching fraction is more than twice as large for the decay into 
$\tilde \chi_2^0$ than for decays to the LSP. The reverse is true for decays producing a Higgs boson. ($iii$) The $\chi_3^0$ decays mediated via the $Z$-boson and the Higgs, to either the LSP or to $\tilde \chi_2^0$, are seen 
to have similar branching fractions.  Clearly it is not advantageous to completely suppress the Higgs mode, which can only be accomplished by reducing $\Delta m_{31}$ to values below the Higgs mass.  Interestingly, this scenario would then also 
predict a signal in the $h+$MET channel. We obtain similar results for the other squark parent scenarios.   

\begin{figure}[htbp]
\centering
\includegraphics[width=0.45\textwidth]{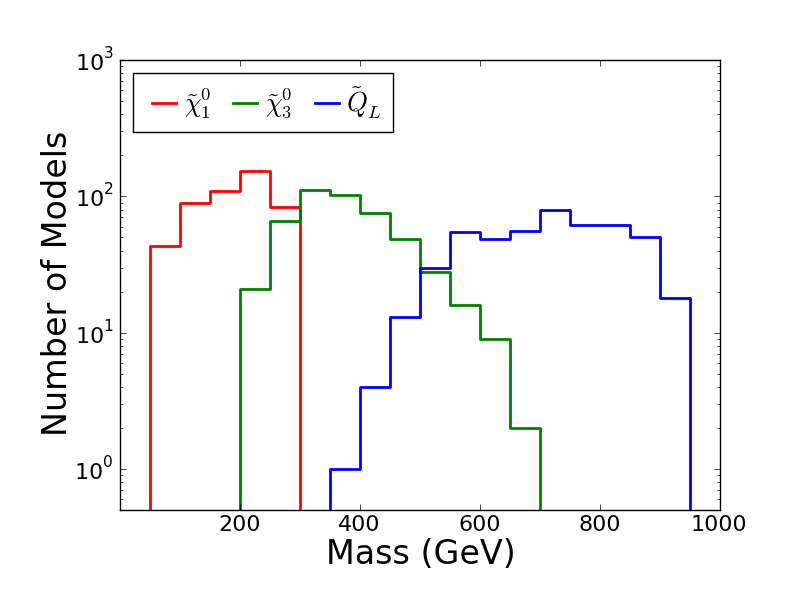}
\includegraphics[width=0.45\textwidth]{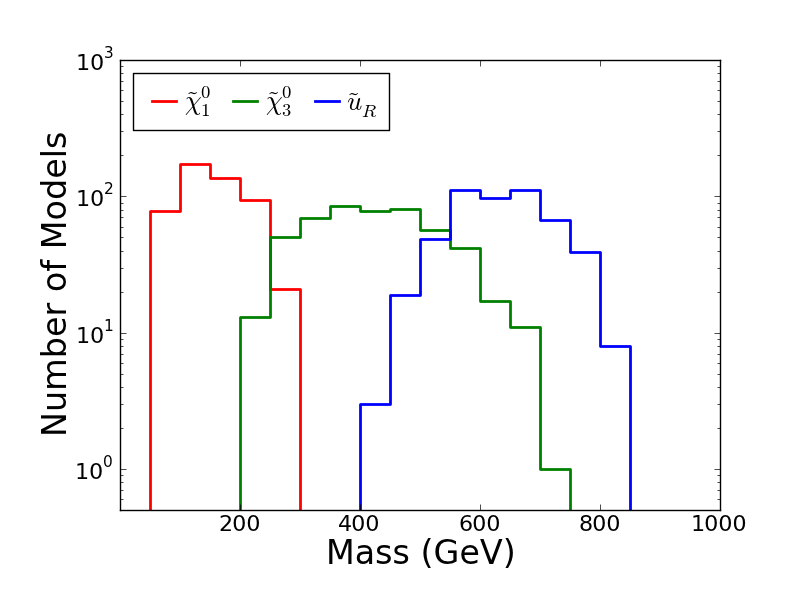}
\includegraphics[width=0.45\textwidth]{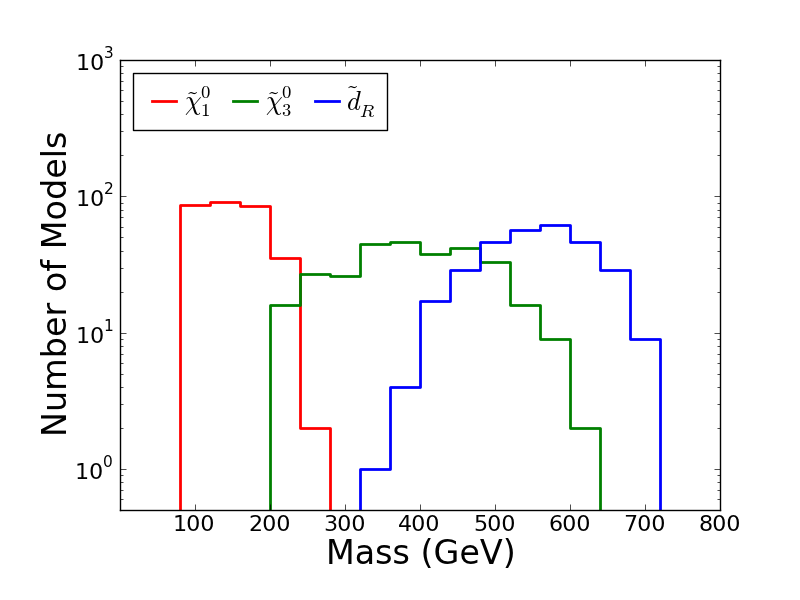}
\caption{Mass distributions of the parent squark, $\chi_3^0$ and $\chi_1^0$ states for the models from our grid scan that agree with all null search results and yield at least
5 events in the ATLAS 20 fb$^{-1}$ $Z+$MET channel.  The top-left, top-right, and bottom panels correspond to the parent squark cases of $\tilde Q_L$,
$\tilde u_R$, and $\tilde d_R$, respectively.}
\label{massdis_histos}
\end{figure}

Lastly, it is interesting to examine the mass distributions of the squarks, bino-like $\tilde \chi_3^0$ and Higgsino-like LSP states 
in the models that successfully reproduce the ATLAS 20 fb$^{-1}$ $Z+$MET signal (here defined to be $N \geq 5$ $Z$+MET events).  This is presented for the three parent squark types in 
Fig.~\ref{massdis_histos}. Here we see that the overall spectra of these three sparticles gradually become lighter as we compare 
the $\tilde Q_L$ parent to $\tilde u_R$ and then to $\tilde d_R$, reflecting the corresponding falling squark pair 
production cross sections, with these distributions peaking at 750, 650, and 600 GeV, respectively. In all three cases the peak 
of the $\chi_3^0$ distribution is near $\sim 350$ GeV resulting in a softening of the jets on average, for the $\tilde u_R/\tilde d_R$ cases compared
to the $\tilde Q_L$ parent squark, due to a 
compression of the spectrum.  
The peak of the LSP mass distribution lies roughly near $\sim 200$ GeV for the $\tilde Q_L$ parent 
squark and near $\sim 150$ GeV for both the $\tilde u_R$ and $\tilde d_R$ cases, implying that $\Delta m_{31}$ prefers to lie 
near $\sim 150-200$ GeV in all cases.

\section{Conclusions}

We have examined the $3\sigma$ $Z+$MET excess observed by the ATLAS collaboration in Run I of the LHC in the context of a
Supersymmetric framework.  We have employed the freedom inherent in the pMSSM parameter space to explore whether Supersymmetric 
models can be constructed that generate the observed excess, while simultaneously being consistent with the numerous other null SUSY searches
at the LHC.  Using a large pMSSM model sample that we had previously generated, we found a handful of points that satisfied our critieria,
demonstrating the power of this approach.  
These points shared charateristics for the sparticle spectrum that are crucial for describing the data, namely relatively light $1^{st}/2^{nd}$
generation squarks that decay into a bino-like neutralino, which in turn decays into a light Higgsino multiplet, {\it i.e.}, $\tilde q\to\tilde B\to\tilde h$.
Using these points as seeds, we performed three grid scans, corresponding to the possible types of the parent squark, $\tilde Q_L$,
$\tilde u_R$ and $\tilde d_R$.  We scanned over the set of relevant parameters, $m_{\tilde q}$, $M_1$ and $\mu$,  
and generated three sets of simplified pMSSM models within a limited kinematic range.  All other strongly-interacting sparticles were
set to the same value as in the parent pMSSM model ($\sim 2$ TeV) and $A_t$ was varied to reproduce the observed Higgs boson mass.  
In principle it is possible that light stop and sbottom squarks could also contribute to the signal, but we
limited our analysis here to the simplest scenario.

We then examined the properties of these simplified models in detail.
They predict a range of event rates, up to 21 events, for the ATLAS 20 fb$^{-1}$ $Z+$MET channel, in agreement with the ATLAS measurement.  Several
hundred of our grid points were found to be consistent with the 95\% C.L. bounds from all simulated searches, including the ATLAS 0l,1l+jet and CMS $Z+$MET search
channels.  The case of a left-handed doublet parent squark, $\tilde Q_L$, is found to yield the best fit to the data, with the other scenarios giving slightly smaller event rates.  The sweet spot for the sparticle spectrum is found to have squark masses in the 500-750 GeV range, with
bino masses near 350 GeV with a mass splitting of $150-200$ GeV with the Higgsino LSP.   The bino $\chi_3^0$ state has important decays
involving $W$ and Higgs bosons, as well as the $Z$-boson.  The predicted event rates for these models are close to the 95\% C.L. limits
from the 0l+jets search and the CMS $Z+$MET analysis, but lie somewhat further from the 1l+jet search results.

In conclusion, we have constructed a simplified Supersymmetric model based on the pMSSM, with specific characteristic features 
that successfully yields an
excess for the 20 fb$^{-1}$ $Z+$MET ATLAS analysis, while evading all other SUSY searches at the Run I LHC.  The operations at the 13 TeV LHC
currently underway will be able to quickly discover, or exclude, this scenario.  If the ATLAS $3\sigma$ excess is confirmed with the new data
set, it could very well be a signal for Supersymmetry.

\section{Acknowledgments}

The authors would like to thank Brian Petersen for communications. 
This work was supported by the Department of Energy, Contracts DE-AC02-06CH11357, DE-AC02-76SF00515 and DE-FG02-12ER41811.

\end{document}